\documentstyle[prl,aps,epsfig,multicol]{revtex}
\begin{document}
\def\be{\begin{equation}}
\def\ee{\end{equation}}
\def\bearr{\begin{eqnarray}}
\def\eearr{\end{eqnarray}}
\def\l{\left}
\def\r{\right}

\draft
\title{Vortex formation in a slowly rotating Bose-Einstein condensate
confined in a harmonic-plus-gaussian laser trap}

\author{Tarun Kanti Ghosh}
\address
{ The Abdus Salam International Centre for Theoretical
Physics, Strada Costiera 11, 34014 Trieste, Italy.}

\date{\today}
\maketitle

\begin{abstract}
Motivated by the recent experiment at ENS [ V. Bretin, S. Stock, Y. Seurin and, J. Dalibard, 
Phys. Rev. Lett. {\bf 92}, 050403 (2004)], we study a rotating (non-)interacting
atomic Bose-Einstein condensate confined in a harmonic-plus-gaussian laser trap potential. 
By adjusting the amplitude of the gaussian laser potential, one can make quadratic-plus-quartic potential,
purely quartic potential, and quartic-minus-quadratic potential.  
We show that an interacting Bose-Einstein condensate confined in a harmonic-plus-gaussian laser trap
breaks the rotational symmetry of the Hamiltonian when rotational frequency is greater than 
one-half of the lowest energy surface mode frequency. 
We also show that by increasing the amplitude of the gaussian laser trap, a vortex  appears 
in a slowly rotating Bose-Einstein condensate. Moreover, one can also create a vortex 
in a slowly rotating non-interacting Bose-Einstein condensate confined in harmonic-plus-gaussian laser 
potential.
\end{abstract}
\pacs{PACS numbers: 03.75.Lm, 05.30.Jp}   

\begin{multicols}{2}[]
The rotation of a macroscopic quantum fluid exhibits interesting 
counter-intuitive phenomena. For example, an atomic interacting(repulsive) 
Bose-Einstein condensate (BEC) confined in a rotating harmonic trap produces 
quantized vortices for a sufficiently large rotation \cite{mad,mad1,abo}. 
The theoretical calculation for $ \Omega_c $ based on purely thermodynamic
arguments is significantly smaller than the observed value of 
$\Omega_c $ which is the order of $ \sim 0.7 \omega_0 $, where $\omega_0 $ is the radial 
trap frequency. 
The vortex formation of a harmonically trapped atomic BEC is related to the dynamical 
instability of the lowest energy surface mode excitations whose energy scale is set 
by the harmonic potential \cite{castin}. Above a certain rotation frequency of a harmonically 
symmetric trapped interacting BEC, the system itself starts deforming from a 
circular shape to an elliptic shape and hence it breaks the original rotational symmetry of 
the Hamiltonian. This has been predicted only for a harmonically trapped BEC \cite{recati} 
and detected at the ENS experiment which leads to the vortex nucleation \cite{madison}.
The vortex nucleation starts when the average angular momentum of each particle 
is one which has been measured experimentally \cite{mad1}.

The harmonically trapped BEC becomes singular when the rotation frequency
is equal or greater than the harmonic trap frequency, because the outward centrifugal force 
counteracts the inward force from the harmonic trap. 
One can eliminate the singularity at $ \Omega \geq 1 $ by considering an additional stiffer radial 
potential (say, quartic potential for simplicity).
There is a growing interest about the effect of an anharmonic potential on the properties 
of a rotating BEC \cite{fetter,lundh,ueda,baym,lundh1,tarun,aftalion}.
In spite of the fact that the $ \Omega_c $ in a harmonically trapped BEC
calculated from the thermodynamic arguments does not match with the
experimental values, many authors have studied $ \Omega_c $ in a BEC confined in
quadratic-plus-quartic trap based on the thermodynamic arguments \cite{lundh,ueda,baym,lundh1}.
In this work, we show that a vortex appears in a BEC confined in a quadratic-plus-gaussian laser potential
due to the spontaneous shape deformation of the system as it is happens in a BEC confined by
harmonic trap only and calculate the correct critical rotational frequency to create a single vortex. 
We also show that by increasing the magnitude of the laser trap, a vortex can nucleate in a slowly
rotating Bose gas.

Recently, the quadratic-plus-quartic potential has been achieved experimentally
by superimposing a blue detuned laser beam to the magnetic trap holding the atoms \cite{fast}.
The effective external potential is
\be
V_t(r, z) = \frac{1}{2}m \omega_0^2 \l [r^2 + \epsilon (x^2-y^2) + \frac{\omega_z^2}{\omega_0^2} z^2 
\r ] + U(r),
\ee
where $ \epsilon = \frac{\omega_x^2 - \omega_y^2}{\omega_x^2 + \omega_y^2} $ is the deformation
parameter due to the stirring potential which rotates the system and creates an anisotropic 
potential in the $xy$ plane. 
Also, $2 \omega_0^2 = \omega_x^2 + \omega_y^2 $ and the potential created by the gaussian laser 
beam is  
\be
U(r) = U_0 e^{-\frac{2r^2}{w^2}}.
\ee 
At the ENS experimental set up \cite{fast}, the laser waist is $ w = 25 \mu m$ and the
amplitude of the laser beam is $ U_0 \sim 1.242 \times 10^{-30} J $. The harmonic trap frequencies
are $ \omega_0 = 2\pi \times 75.5 Hz $ and $ \omega_z = 2\pi \times 11 Hz $.
Typically, the size of a condensate is the order of few $\mu m$. 
Since $ \frac{\sqrt 2 r}{w} < 1 $, one can expand the gaussian laser potential,
\be
U(r) = U_0 e^{-\frac{2r^2}{w^2}} \sim U_0(1 - \frac{2r^2}{w^2} + \frac{2r^4}{w^4}).
\ee 
The resulting potential can be written as,
$$
V_t(\tilde r) = \frac{\hbar \omega_0}{2} \l [(1-k) \tilde r^2 + \epsilon \l (\tilde x^2 - 
\tilde y^2 \r ) + \l (\frac{\omega_z^2}{\omega_0^2} \r ) \tilde z^2 + \lambda \tilde r^4 \r ],
$$ 
where  the co-ordinates $ \tilde r, \tilde x, \tilde y $, and $ \tilde z $ are in units of the harmonic 
oscillator
length $ a_0 = \sqrt{\frac{\hbar}{m \omega_0}} = 1.475 \mu m $. 
Here, $ k = \frac{4 U_0}{m 
\omega_0^2 w^2} $ and $ \lambda = k (\frac{a_0}{w})^2 $. For the given values of the $ w,  U_0$
and $ \omega_0 = 2 \pi \times 75.5 Hz $, $ k = 0.24 $ and $ \lambda =  8.297 \times 10^{-4} $
at the ENS experiment \cite{fast}.
The value of $k$ renormalize the oscillator frequency ($\omega_0$) 
of the magnetic trap and also the anharmonic term $\lambda $ depends on the value of $k$.
At the ENS experiment, $k$ is small. In this paper, we are interested to study when $k$ is large.
When $k <1 $, $ k =1 $ and $ k>1 $, the effective potential is quadratic-plus-quartic potential,
only quartic potential and quartic-minus-quadratic potential, respectively.
If $ k > 1$, the effective external potential becomes Mexican hat structure. In principle,
$ k \geq 1$ can be easily obtain in the current experimental set up at ENS by increasing $U_0$ from 
$U_0 = 1.242 \times 10^{-30} J $ to $U_0 \geq 5 \times 10^{-30} J$. 
For simplicity, we study quasi-2D system since the elongated condensate expands radially and contracts
axially when it rotates rapidly.
The equation of motion of the condensate wave function $\psi (\vec r)$ is described by the
mean-field Gross-Pitaevskii equation,
$$
i \hbar \frac{ \partial \psi (\vec r)}{\partial t} = \l [ - \frac{\hbar^2}{2m} \nabla^2
+ V_t(\vec r) + g_{2} |\psi(\vec r)|^2 - \Omega_0 L_z \r ] \psi(\vec r).
$$
Here, $ g_2 = 2 \sqrt{2\pi}\hbar \omega_z a a_z $ \cite{ho} is the strength of the mean-field 
interaction, $a $ is the $s$-wave scattering length and $a_z =\sqrt{\frac{\hbar}{m \omega_z}} $. 
Also, $ L_z = x p_y - y p_x $ is the $z$-component of the angular momentum operator 
and $ \Omega_0 $ is the trap rotation frequency. When $ \epsilon = 0 $, it preserves the circular symmetry
of the system.

One can write down the Lagrangian density corresponding to the quasi-2D system as follows:
\bearr \label{density}
{\cal L} & = & \frac{ i \hbar }{ 2 } \l (\psi\frac{\partial{\psi^{*}}}{\partial{ t }} -
\psi^{*}\frac{\partial\psi}{\partial t} \r ) \\ \nonumber & + &
\l (\frac{\hbar^{2}}{2m} |\nabla \psi
|^{2} + V_t({x, y})|\psi |^{2} + \frac{g_2}{2} |\psi |^{4} - \Omega \psi^{*} L_z \psi \r ).
\eearr

Here, we use the time-dependent variational method to study the properties
of a rotating BEC confined in a quadratic-plus-laser potential.
In order to obtain the evolution of the condensate
we assume the most general Gaussian wave function,
\be \label{wave}
 \psi \l (\tilde x, \tilde y, t \r ) = C(t) e^{-\frac{1}{2} \l [ \alpha (t) \tilde x^2 + \beta (t) \tilde y^2 
- 2 \gamma (t) \tilde x \tilde y \r ]},
\ee
 where $ C(t) $ is the normalization constant. 
Further, $ \alpha(t) = \alpha_1(t) + i \alpha_2(t) $,
$ \beta(t) = \beta_1(t) + i \beta_2(t) $ and $ \gamma(t) = \gamma_1(t) + i \gamma_2(t) $ are the 
time-dependent dimensionless complex variational parameters. The $ \alpha_1 $ and $ \beta_1 $ are
inverse square of the condensate widths in $ x $ and $ y $ directions,
respectively. 
The above mentioned order parameter describes only the vortex free
condensate (irrotational system) as the phase ($ S(x,y) = \gamma xy $) 
corresponds to the irrotational velocity flow.

We obtain the variational Lagrangian  $ L $ by substituting Eq. (\ref{wave})
into Eq. (\ref{density}) and integrating the Lagrangian density over the
space co-ordinates,
\bearr
\frac{L}{N\hbar\omega_{0}} & = & \nonumber
\frac{1}{4 D } [-(\beta_{1}\dot{\alpha_{2}}+\alpha_{1}\dot{\beta_{2}} - 2    
\gamma_1 \dot{\gamma_2}) + (\alpha_{1} + \beta_{1} ) D  
\\ \nonumber & + &   (\alpha_{2}^{2} + \gamma_{2}^2 ) \beta_{1}
 +  ( \beta_{2}^2 + \gamma_{2}^2 ) \alpha_{1} \\ \nonumber & - &  2 (
\alpha_{2} + \beta_{2} ) \gamma_{1} \gamma_{2}
 + (1-k)(\alpha_1 + \beta_{1}) + \epsilon( \beta_1 - \alpha_1)
\\ \nonumber & + & \frac{\lambda}{2} \frac{(3 \alpha_1^2 
+ 2 \alpha_1 \beta_1 + 3 \beta_1^2 + 4 \gamma_1^2)}{D} 
\\ & + & P D^{3/2}
 + 2 \Omega (\gamma_1(\beta_2 -\alpha_2)+ \gamma_2(\alpha_1
-\beta_1))],
\eearr
where $ D = \sqrt{ \alpha_1 \beta_1 - \gamma_1^2} $ and $ P = 2 \sqrt{\frac{2}{\pi}} \frac{(N-1) a}{a_{z}} $ 
is 
the dimensionless parameter
that indicates the strength of the mean-field interaction and it can be positive or negative depending on 
the sign of the $s$-wave scattering length $a$ \cite{fesh}.

The variational energy of the rotating condensate at equilibrium is given  in
terms of the inverse square width of the condensate
along the $x$ and $y$ directions and the phase parameter $ \delta \gamma_2$,
\bearr
\frac{ E}{N\hbar \omega_0} &  = & \nonumber \frac{1}{4} [ (\alpha_1 +
\beta_1  ) + (\gamma_2^2 + 1- k) (\frac{1}{\alpha_1 } + \frac{1}{\beta_1 })
\\ \nonumber & + & \epsilon (\frac{1}{\alpha_1 } - \frac{1}{\beta_1 })
+\frac{\lambda}{2}(\frac{3}{\alpha_1^2}+ \frac{2}{\alpha_1 \beta_1} +\frac{3}{\beta_1^2})
\\  & + &  P \sqrt{\alpha_1  \beta_1 } + 2 \Omega \gamma_2
(\frac{1}{ \beta_1 } - \frac{1}{\alpha_1 })],
\eearr
where $ \Omega = \Omega_0/ \omega_0 $.
We have used the fact that $ \alpha_2 = \beta_2 = \gamma_1 =0 $ to obtain the above equation.

One can get the equilibrium value of the variational parameters, $
\alpha_{10} = X $, $ \beta_{10} = Y $ and $ \gamma_{20} = Z $ by minimizing the energy
with respect to the variational parameters,

\be \label{c1}
X^2(1 + \frac{P}{2} \sqrt{\frac{Y}{X}}) =
Z^2 - 2 \Omega Z + 1 - k + \epsilon + \lambda (\frac{3}{X}+ 
\frac{1}{Y}),
\ee

\be \label{c2}
Y^2(1 + \frac{P}{2}  \sqrt{\frac{X}{Y}}) =
Z^2 + 2 \Omega Z + 1 - k - \epsilon +  \lambda (\frac{1}{X}+ 
\frac{3}{Y}),
\ee

and
\be \label {c3}
Z = \Omega \frac{(Y - X)}{(Y + X)}.
\ee
Eqs. (\ref{c1}), (\ref{c2}) and (\ref{c3}) describes how the shape of the condensate
changes due to the change of the amplitude of the laser beam as well as external rotation.

The average angular momentum per particle is given by,
\be \label{ang}
\frac{<L_z>}{N \hbar} = \frac{1}{2}  Z (\frac{1}{Y} -
\frac{1}{X}) = \frac{\Omega}{2}\frac{ (1-\eta)^2}{\eta (X +
Y)},
\ee
where $\eta = \frac{Y}{X}$ is the ratio of the square of the widths along the $x$ and $y$ directions.
Eq. (\ref{ang}) explicitly shows how the angular momentum is transfered to the trapped
BEC due to the shape deformation with the rotation frequency.
The vortex would appear when $ <L_z> = N \hbar $.
Therefore, one can estimate the critical angular frequency to create a single vortex
from the relation: $ <L_z> = N \hbar $.

First we calculate the quadrupole mode frequency of BEC confined in a magnetic-plus-laser potential.
We expand the Lagrangian in the following way: $ \alpha = X + \delta \alpha_1 $,
$ \beta = Y + \delta \beta_1 $, and $ \gamma = \delta \gamma_1 $. We keep only the
second order deviations from their equilibrium values.
Then we calculate the Lagrangian quadratic in the deviations
and using the Euler-Lagrangian equation of motion, we get the following two coupled equations 
of $ \delta \alpha_1 $ and $ \delta \beta_1 $:

\bearr
\delta \ddot \alpha_1 & = & \nonumber [\frac{P}{2} X^2 \sqrt{ \frac{Y}{X }} 
- 4(1 - k + \epsilon) - 2 \lambda ( \frac{9}{X} +  \frac{2}{Y})] \delta \alpha_1
\\ & - & [\frac{P}{2} X^2 \sqrt{\frac{X}{Y}} + 2 \lambda 
\frac{X}{Y^2}] \delta \beta_1,
\eearr

\bearr
\delta \ddot \beta_1 & = & - [\frac{P}{2} Y^2 \sqrt{\frac{Y}{X}} + 2 \lambda
\frac{Y}{X^2}] \delta \alpha_1 
\\ \nonumber & + &  [\frac{P}{2}Y^2 \sqrt{\frac{X}{Y}} - 4(1 - k - \epsilon) - 
2 \lambda ( \frac{2}{X} +  \frac{9}{Y})] \delta \beta_1.
\eearr

For an isotropic trap, $ \epsilon = 0 $, and we set $ \delta \alpha_1 = - \delta \beta_1 $ to 
calculate the lowest energy quadrupole mode frequency which is given by,
\be
\omega_q^2 =  4(1-k) + \frac{20 \lambda}{R_0} - P R_0^2,
\ee
where $ R_0 $ is the equilibrium radius of the system, and this  can be obtained
from the real solution of the cubic equation: $ (1+\frac{P}{2}) R_0^3 - (1-k) R_0 - 4 \lambda = 0 $.
The above quadrupole frequency $ \omega_q $ is valid for all interaction (repulsive, attractive) 
strength.
The quadrupole mode frequency $ \omega_q $ vs $ P $ for different values of $k$, i.e. for 
different configuration of the effective potential, is shown in Fig.1.
\begin{figure}[h]
\epsfxsize 9cm
\centerline {\epsfbox{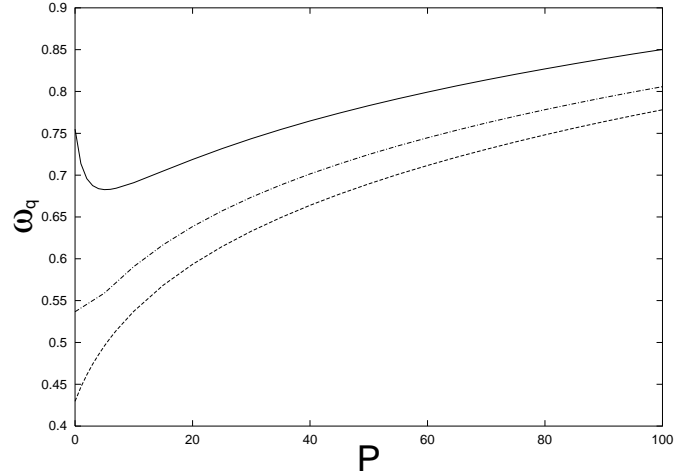}}
\vspace{0.2 cm}
\caption{Plots of quadrupole mode frequency $\omega_q $ vs interaction parameter $P$ for $ k = 0.9,
\lambda = 0.0031 $(upper curve), $ k = 1.0, \lambda = 0.0034 $(middle curve) and 
$ k = 1.1, \lambda = 0.0038 $ (lower curve).}
\end{figure}
Fig.1 shows that when $ k<1$, the $ \omega_q $ decreases as $P $ increases and after
a critical value of $P$, $ \omega_q $ start increasing slowly. On the other hands, for $ k\geq 1$,
$ \omega_q $ increases as $P$ increases.
\begin{figure}[h]
\epsfxsize 9cm
\centerline {\epsfbox{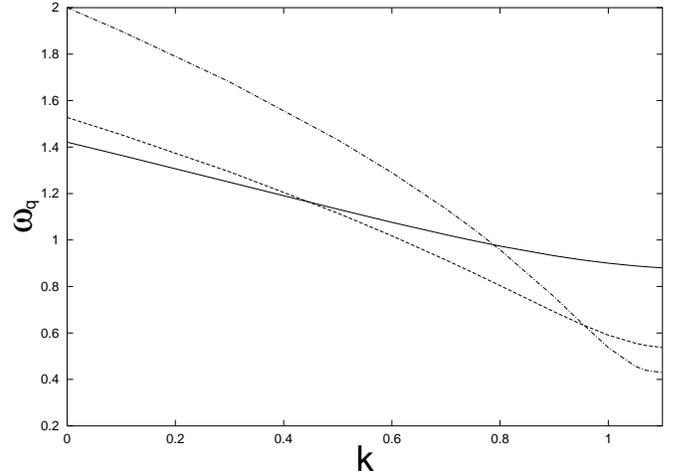}}
\vspace{0.2 cm}
\caption{Plots of quadrupole mode frequency $\omega_q $ vs $k$ for $ P = 0 $(dot-dashed), 10 (dashed), 200
(solid).}
\end{figure}
The quadrupole mode frequency $\omega_q $ vs $k$ for different values of $P$ is shown in
Fig.2.  For $ k=0 =P $, it reproduces the known result for the quadrupole mode frequency of a 
harmonically trapped noninteracting BEC. 
With the increasing of $k$, $\omega_q $ of an ideal Bose gases is decreasing
faster than the interacting cases.
 
The centrifugal term $ - \Omega L_z $ shifts the quadrupole mode frequency by $ -2 \Omega $.
The quadrupole mode frequency with $ m_z = -2 $ is 
\be
\omega_{-2} = \sqrt{ 4(1-k) + \frac{20 \lambda}{R_0} - P R_0^2} - 2 \Omega.
\ee

We will discuss the effect of the harmonic-plus-gaussian laser trap potential on the possibility of a 
vortex formation in a slowly rotating BEC.
Using the three coupled equations (\ref{c1}), 
(\ref{c2}) and (\ref{c3}), we get the following two coupled polynomial equations:

\bearr \label{c11}
0 & = & \nonumber (X^3 Y^3 + 2 X^4 Y^2 + X^5 Y)(1+\frac{P}{2} 
\sqrt{\frac{Y}{X}}) \\ \nonumber & - & X^3 Y ( 3 \Omega^2 + 1 - k + 
\epsilon ) + (X Y^3 + 2 X^2 Y^2) \times 
\\ & &(\Omega^2 - 1 + k - \epsilon) -  \lambda (5 X^2 Y + 7 X Y^2  + X^3 + 3 Y^3)
\eearr
 and
\bearr \label{c22}
 0 & = & \nonumber (X^3 Y^3 + 2 X^2 Y^4 + X
Y^5)(1+\frac{P}{2}
\sqrt{\frac{X}{Y}}) \\ \nonumber & - &  X Y^3 ( 3 \Omega^2 + 1 - k +
\epsilon ) + (X^3 Y + 2 X^2 Y^2) \times 
\\ & & (\Omega^2 - 1 + k - \epsilon) - \lambda (5 X Y^2 + 7 X^2 Y + 3 X^3 + 
Y^3).
\eearr
For simplicity, we put $ \epsilon = 0 $ and solve those two coupled equations
numerically. 
We find that spontaneous deformation occurs when the rotational frequency is 
greater than one-half of the quadrupole mode frequency $ \omega_{q} $, and start
transferring the angular momentum to the system [see Eq. (\ref{ang})]. 
This circular symmetry breaking is due
to the tendency against an instability of the lowest energy surface mode frequency
with $m_z = - 2 $.    
The variation of $\eta $ with the rotational frequency $\Omega $ for $\epsilon =0 $
and $ k = 0.9 $ is shown in Fig. 3. The same analysis can also be done for other 
values of $k$ and $P$.
\begin{figure}[h]
\epsfxsize 9cm
\centerline {\epsfbox{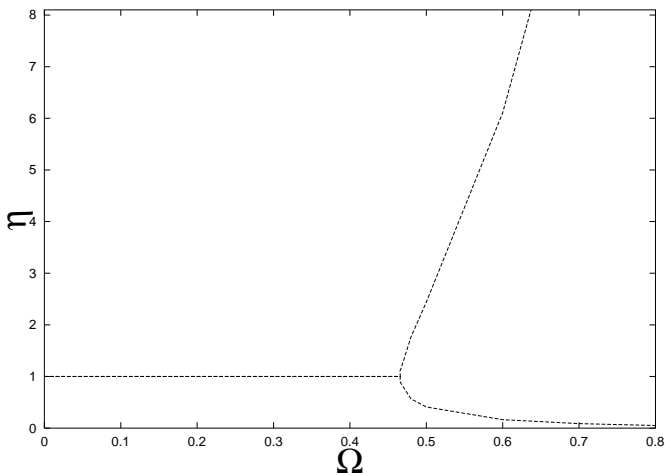}}
\vspace{0.2 cm}
\caption{The variation of $\eta $ vs the rotational frequency $\Omega$ for
$P=200$, $k = 0.9 $, and $\epsilon = 0$.}
\end{figure}
For $k=0$ (only harmonic potential) and $ P $ is large, one can produce a vortex when $\Omega \sim 0.7$.
For $ k \neq 0 $, the critical rotational frequency ($\Omega_c $) to create a single vortex 
is small ($\Omega_c < 0.7 $) compared to harmonic trap case ($\Omega_c \sim 0.7 $). By adjusting the 
strength of the gaussian laser beam, one can produce a vortex in a slowly rotating BEC.   
For a purely harmonic trapped ideal Bose gases, the lowest energy surface mode frequency
is $ 2 \omega_0$ and hence instability will occur when $ \Omega = \omega_0$ at which the
system is destabilized. Therefore, there is no spontaneous shape deformation and hence
a vortex can not appear in the noninteracting gas confined by harmonic trap only.
When $ k\neq 0 $, the the lowest energy surface mode frequency becomes less than $ 2 \omega_0$
and an instability will occur when $ \Omega < \omega_0$ at which the system is stabilized. 
There is a spontaneous shape deformation and one can produce a vortex in a slowly rotating 
noninteracting Bose gases confined in a harmonic-plus-gaussian laser trap. 
\begin{figure}[h]
\epsfxsize 9cm
\centerline {\epsfbox{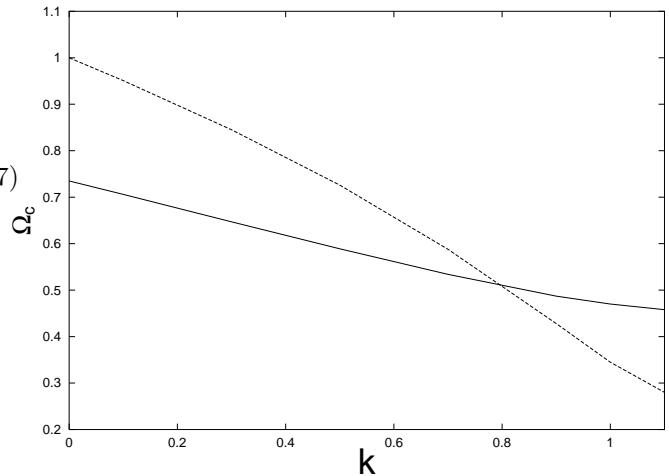}}
\vspace{0.2 cm}
\caption{Critical rotational frequency $\Omega_c $ vs $k$ for 
$ P = 200 $(solid line) and $ P = 0 $ (dashed line).}
\end{figure}
Fig. 4 shows how $ \Omega_c $ is decreasing with the parameter $k$ for 
different values of $P$. 
As $k$ increases, $ \Omega_c $ is also decreasing.
Hence there is a possibility of vortex formation of a slowly rotating BEC if $k$ is finite.  
When $ k<0.8$, $ \Omega_c $ is small in the Thomas-Fermi regime compared to
the non-interacting case. When $ k> 0.8$, $ \Omega_c $ is large in the Thomas-Fermi regime
compared to the non-interacting case.  
It implies that the two-body interaction helps to create a vortex when $ k < 0.8$, but the two-body
interaction do not help when $ k > 0.8$.

In this work, we have shown that
by adjusting amplitude of the gaussian laser beam one can make quadratic-plus-quartic potential, purely 
quartic
potential and quartic-minus-quadratic potential.
We have also shown that an interacting BEC confined in a harmonic-plus-gaussian laser tarp
breaks the rotational symmetry of the Hamiltonian due to the instability of the lowest energy 
surface mode. It is argued that by increasing the amplitude of the gaussian laser trap, a vortex 
can be created in a slowly rotating interacting as well as noninteracting BECs.

\end{multicols}
\end{document}